\documentclass[journal=jctcce,manuscript=article,layout=twocolumn]{achemso}
\usepackage[utf8]{inputenc}
\usepackage[T1]{fontenc}
\usepackage{lmodern}

\usepackage{amsmath}
\usepackage{amsfonts}
\usepackage{amssymb}
\usepackage{fixmath}
\usepackage{microtype}
\usepackage{braket}
\usepackage{multirow}
\newcommand{\specialcell}[2][c]{ %
  \begin{tabular}[#1]{@{}c@{}}#2\end{tabular}}

\newcommand*{\abs}[1]{\left|#1\right|}
\newcommand*{\pvec}[1]{\vec{#1}\mkern2mu\vphantom{#1}'}
\usepackage{bm}\newcommand*{\mat}[1]{\bm{#1}}

\newcommand*{\dd}{\mathrm{d}}

\author{Robert R\"uger}
\affiliation[Scientific Computing \& Modelling]{Scientific Computing \& Modelling NV, De Boelelaan 1083, 1081 HV Amsterdam, The Netherlands}
\alsoaffiliation[VU University Amsterdam]{Department of Theoretical Chemistry, VU University Amsterdam, De Boelelaan 1083, 1081 HV Amsterdam, The Netherlands}
\author{Erik van Lenthe}
\affiliation[Scientific Computing \& Modelling]{Scientific Computing \& Modelling NV, De Boelelaan 1083, 1081 HV Amsterdam, The Netherlands}
\author{You Lu}
\affiliation[Scientific Computing \& Modelling]{Scientific Computing \& Modelling NV, De Boelelaan 1083, 1081 HV Amsterdam, The Netherlands}
\altaffiliation{Present address: STFC Daresbury Laboratory, Daresbury, Warrington WA4 4AD, United Kingdom}
\author{Johannes Frenzel}
\affiliation[University of Calgary]{Department of Chemistry, University of Calgary, 2500 University Drive, N.W., T2N 1N4 Calgary, Canada}
\altaffiliation{Present address: Lehrstuhl für Theoretische Chemie, Ruhr-Universit\"at Bochum, Universitätsstraße 150, 44780 Bochum, Germany}
\author{Thomas Heine}
\affiliation[Jacobs University Bremen]{School of Engineering and Science, Jacobs University Bremen, Campus Ring 1, 28759 Bremen, Germany\looseness -1}
\author{Lucas Visscher}
\email{l.visscher@vu.nl}
\affiliation[VU University Amsterdam]{Department of Theoretical Chemistry, VU University Amsterdam, De Boelelaan 1083, 1081 HV Amsterdam, The Netherlands}

\title[Intensity-Selected TD-DFTB]{Efficient Calculation of Electronic Absorption Spectra by Means of Intensity-Selected TD-DFTB\looseness -1}

\abbreviations{IR,NMR,UV}

\begin{document}







\begin{abstract}
During the last two decades density functional based linear response approaches have become the de facto standard for the calculation of optical properties of small and medium-sized molecules. At the heart of these methods is the solution of an eigenvalue equation in the space of single-orbital transitions, whose quickly increasing number makes such calculations costly if not infeasible for larger molecules. This is especially true for time-dependent density functional tight binding (TD-DFTB), where the evaluation of the matrix elements is inexpensive. For the relatively large systems that can be studied the solution of the eigenvalue equation therefore determines the cost of the calculation. We propose to do an oscillator strength based truncation of the single-orbital transition space to reduce the computational effort of TD-DFTB based absorption spectra calculations. We show that even a sizeable truncation does not destroy the principal features of the absorption spectrum, while naturally avoiding the unnecessary calculation of excitations with small oscillator strengths. We argue that the reduced computational cost of intensity-selected TD-DFTB together with its ease of use compared to other methods lowers the barrier of performing optical properties calculations of large molecules, and can serve to make such calculations possible in a wider array of applications.
\end{abstract}




\section{Introduction}

Density functional theory (DFT) based on the Hohenberg-Kohn theorem~\cite{HohenbergKohnTheorem1964} and implemented in terms of the Kohn-Sham equations~\cite{KohnShamEquations1965} is one of the most popular methods in both solid-state physics and quantum chemistry. The reason for this popularity is that DFT is computationally relatively affordable and its accuracy for many systems not far behind more accurate but also much more expensive wavefunction based methods. For systems which are too large to be treated with DFT one can introduce further approximations on top of the DFT framework, most notably density functional based tight binding (DFTB)~\cite{PorezagDFTB1995,SeifertDFTB1996}. In DFTB, tight-binding approximations are made to the DFT total energy expression, most notably an optimized minimum valence orbital basis that reduces the linear algebra operations, and a two center-approximation that allows to precalculate and store all integrals using the Slater-Koster technique~\cite{SlaterKoster1954}. The self-consistent charge (SCC) technique~\cite{SeifertSCCDFTB1998} accounts for density fluctuations and improves results on polar bonds. Detailed information on the DFTB parameterization for all elements have been published recently~\cite{WahiduzzamanQuasinano2013}.

As the underlying Hohenberg-Kohn theorem is only a statement about the ground state, standard DFT can not be applied to the broad class of problems involving excited states, most notably the study of optical properties of an electronic system. The extension of DFT to excited states has been accomplished in the form of time-dependent density functional theory (TD-DFT) based on the Runge-Gross theorem~\cite{RungeGrossTheorem1984}, which is a time-dependent analogon to the Hohenberg-Kohn theorem. In quantum chemistry TD-DFT is in practice often used in the form of \citeauthor{CasidaTDDFT1995}'s formalism~\cite{CasidaTDDFT1995}, where the electron density's linear response to a perturbation in the external potential is used to construct an eigenvalue equation in the space of single orbital transitions from which the excitation energies and excited states can be extracted. TD-DFT calculations of excited states are much more expensive than their ground state counterpart, and therefore limited in the size of the systems that can be treated. At the expense of accuracy the computational cost of TD-DFT calculations can be reduced by making further approximations, most notably the Tamm-Dancoff approximation\cite{HirataTDA1999} (TDA) and related techniques\cite{GrimmeSimplifiedTDDFT2013}. It is interesting to note that TDA results can even be better than unapproximated TD-DFT results\cite{CasidaTDDFTReview2012} even though TDA violates the Thomas-Reiche-Kuhn $f$-sum rule~\cite{Thomas1TRKFSum1925,Kuhn2TRKFSum1925,ReicheThomas3TRKFSum1925}. Another way to reduce the computational effort is to translate \citeauthor{CasidaTDDFT1995}'s formalism to the DFTB framework. This was done by \citeauthor{NiehausTDDFTB2001} and is known as time-dependent density functional based tight binding (TD-DFTB)~\cite{NiehausTDDFTB2001}. Note that there is an alternative formulation of TD-DFTB which has recently been developed by \citeauthor{GaussianTDDFTB2011}.\cite{GaussianTDDFTB2011}

At the heart of both TD-DFT and TD-DFTB is the solution of \citeauthor{CasidaTDDFT1995}'s eigenvalue equation in the space of single orbital transitions. As the number of transitions grows quadratically with the size of the system, the resulting matrix can only be diagonalized using iterative eigensolvers, and even then the huge size of the matrix quickly becomes the limiting factor. This is especially true for TD-DFTB where the calculation of the matrix elements is rather cheap, so that bigger systems with relatively larger matrices can be investigated.

In this article we discuss practical methods to deal with the increasing dimension of the eigenvalue problem encountered in TD-DFTB calculations for large molecules. The remainder of the article is organized as follows. In section~\ref{s:Methods} we recapitulate the basic equations of ground state SCC-DFTB and review how adapting \citeauthor{CasidaTDDFT1995}'s TD-DFT approach to the DFTB framework results in the TD-DFTB method. In section~\ref{s:ComputationalMethods} we analyze the bottlenecks of the method and show how the TD-DFTB equations can be implemented efficiently. For the specific application of calculating electronic absorption spectra we present ways to reduce the size of the eigenvalue problem through a physically motivated truncation of the single orbital transition space. In section~\ref{s:Examples} we use this truncation to calculate the absorption spectra of a number of example molecules ranging from small model systems to entire proteins in order to validate the precision of the results as well as the computational performance of the method. Section~\ref{s:Conclusion} summarizes our results.

\section{Review of the methods}\label{s:Methods}

\subsection{DFTB}\label{ss:DFTB}

Let us quickly recapitulate the most important equations of SCC-DFTB~\cite{SeifertSCCDFTB1998}. More comprehensive reviews can be found in reference~\citenum{OliveiraDFTBReview2009} and~\citenum{SeifertDFTBReview2012}. The total energy within the SCC-DFTB method is given by
\begin{align}
\label{eq:DFTBGroundStateEnergy} E_\text{SCC-DFTB} &= E_\text{orb} + E_\text{SCC} + E_\text{rep} \\
E_\text{orb} &= \sum_{i}^{N_\text{occ}} \braket{\phi_i | \hat H^0 | \phi_i} \displaybreak[0]\\
E_\text{SCC} &= \frac{1}{2} \sum_\mathcal{AB}^{N_\text{atom}} \Delta q_\mathcal A \gamma_\mathcal{AB} \Delta q_\mathcal B \\
E_\text{rep} &= \frac{1}{2} \sum_\mathcal{AB}^{N_\text{atom}} U_\mathcal{AB} \; ,
\end{align}
where the individual terms are called the orbital contribution~$E_\text{orb}$, the self-consistent charge correction~$E_\text{SCC}$, and the repulsive energy~$E_\text{rep}$.

DFTB uses a (typically minimal) basis of atomic valence orbitals~$\chi_\mu(\vec r)$ to expand the molecular orbitals~$\phi_i(\vec r)$ as
\begin{equation}\label{eq:OrbitalExpansion}
\phi_i(\vec r) = \sum_{\mathcal A}^{N_\mathrm{atom}} \sum_{\mu \in \mathcal A} c_{\mu i} \chi_\mu(\vec r) \; .
\end{equation}
In this basis the matrix elements of~$\hat H^0$ are calculated as
\begin{equation}
\braket{\chi_\mu | \hat H^0 | \chi_\nu} = \begin{cases}
\varepsilon_\mu^\text{free atom} \;\; \text{for } \mu = \nu \\[12pt]
\Braket{\chi_\mu | \hat T + \hat V^0_\mathcal{AB} | \chi_\nu} \\ \hfill\text{ for } \mu \in \mathcal A, \; \nu \in \mathcal B, \; \mathcal A \neq \mathcal B \\[8pt]
0 \;\; \text{ otherwise}
\end{cases},
\end{equation}
where~$\varepsilon_\mu^\text{free atom}$ is the energy of the corresponding atomic orbital of the free atom and~$\hat V^0_\mathcal{AB}$ is a strictly pairwise effective potential, usually implemented in terms of the Kohn-Sham potential and the atomic electron densities~$\rho^0_\mathcal A$ and~$\rho^0_\mathcal B$. Note that the matrix elements of~$\hat H^0$ only depend on the elements of atom~$\mathcal A$ and~$\mathcal B$ and the distance~$R_\mathcal{AB} = \abs{\vec R_\mathcal A - \vec R_\mathcal B}$ between the two nuclei. It is therefore possible to precalculate them by running DFT calculations for all individual atoms as well as all possible dimers at a sufficient number of internuclear distances~$R_\mathcal{AB}$. Details on this parametrization can be found in the literature~\cite{SeifertSCCDFTB1998,WahiduzzamanQuasinano2013}.

The self-consistent charge contribution~$E_\text{SCC}$ accounts for the fact that the actual ground state density
\begin{equation}
\rho^\text{GS}(\vec r) = \rho^0(\vec r) + \delta \rho(\vec r) \quad \text{with} \quad\hspace{-1.3pt} \rho^0(\vec r) = \sum_\mathcal A \rho_\mathcal A^0(\vec r)
\end{equation}
differs from the sum of the atomic densities by a density fluctuation~$\delta \rho(\vec r)$. Within SCC-DFTB this density fluctuation is then decomposed into atomic contributions~$\delta \rho_\mathcal A(\vec r)$ which are subjected to a multipole expansion and a monopolar approximation.
\begin{equation}
\delta \rho(\vec r) = \sum_\mathcal A \delta \rho_\mathcal A(\vec r) \approx \sum_\mathcal A  \Delta q_\mathcal A \xi_\mathcal A(\vec r)
\end{equation}
Here~$\xi_\mathcal A(\vec r)$ is a spherically symmetric function centered on atom~$\mathcal A$ and the transferred charges~$\Delta q_\mathcal A$ are calculated from the expansion coefficients and the overlap matrix~$S_{\mu\nu} = \braket{\chi_\mu | \chi_\nu}$ through Mulliken population analysis.
\begin{align}\label{eq:DFTBMullikenAnalysis}
\Delta q_\mathcal A &= q_\mathcal A - q_\mathcal A^\text{free atom} \qquad \text{with} \\
q_\mathcal A &= \frac{1}{2} \sum_i^{N_\text{occ}} \sum_{\mu \in \mathcal A} \sum_{\nu} \Big( c_{\mu i} S_{\mu \nu} c_{\nu i} + c_{\nu i} S_{\nu \mu} c_{\mu i} \Big) \nonumber
\end{align}
The elements of the matrix~$\mat \gamma$ in equation~\eqref{eq:DFTBGroundStateEnergy} can now be calculated with any exchange-correlation functional~$E_\mathrm{xc}[\rho]$ through
\begin{gather}
\label{eq:DFTBGroundStateGamma} \gamma_\mathcal{AB} = \int \dd^3 \vec r \int \dd^3 \pvec r \; \xi_\mathcal A(\vec r) \; f_\mathrm{Hxc}[\rho^0](\vec r, \pvec r) \; \xi_\mathcal B(\pvec r) \\[-7pt]
\intertext{with}
f_\mathrm{Hxc}[\rho^0](\vec r, \pvec r) = \frac{1}{\abs{\vec r - \pvec r}} + \frac{\delta^2 E_\mathrm{xc}}{\delta \rho(\vec r) \delta \rho(\pvec r)} \Big|_{\rho^0} \; .
\end{gather}
Note that~$\gamma_\mathcal{AB}$ only depends on the type of atom~$\mathcal A$ and~$\mathcal B$ as well as the distance~$R_\mathcal{AB}$ between their nuclei. Due to the locality of the exchange-correlation functional, the SCC contribution reduces in the limit of large~$R_\mathcal{AB}$ to just the Coulomb interaction between two point charges at~$\vec R_\mathcal A$ and~$\vec R_\mathcal B$. The on-site term~$\gamma_\mathcal{AA}$ can be approximated by the atom's Hubbard parameter
\begin{equation}
U_\mathcal A \approx 2 \eta_\mathcal A \approx I_\mathcal A - A_\mathcal A \; ,
\end{equation}
where~$I_\mathcal A$ is the atomic ionization potential, $A_\mathcal A$ the electron affinity, and~$\eta_\mathcal A$ the chemical hardness which can be calculated by DFT as the second derivative of the energy with respect to the occupation number of the highest occupied atomic orbital. An interpolation formula is then used to calculate~$\gamma_\mathcal{AB}$ for intermediate distances~$R_\mathcal{AB}$.\cite{SeifertSCCDFTB1998}

While the repulsive term~$E_\text{rep}$ can also be parametrized from DFT calculations~\cite{SeifertSCCDFTB1998}, it is for fixed nuclear positions only a global shift in energy that does not influence the absorption spectrum and is hence irrelevant for this article.

Finally the molecular orbitals~$\phi_i(\vec r)$ from equation~\eqref{eq:OrbitalExpansion} can be obtained by solving the Kohn-Sham equation of SCC-DFTB.
\begin{gather}
\label{eq:DFTBKohnShamEquation} \sum_\nu H_{\mu\nu} c_{\nu i} = \varepsilon_i \sum_\nu S_{\mu \nu} c_{\nu i} \pagebreak[0]\\
\begin{split}
H_{\mu\nu} = H^0_{\mu\nu} + \frac{1}{2} S_{\mu\nu} \sum_{\mathcal C} \left( \gamma_\mathcal{AC} + \gamma_\mathcal{BC} \right) \Delta q_\mathcal C \\
\text{with} \quad \mu \in \mathcal A, \nu \in \mathcal B
\end{split}
\end{gather}
Note that this has to be done self-consistently as the~$\Delta q_\mathcal C$ depend on the expansion coefficients via equation~\eqref{eq:DFTBMullikenAnalysis}.

\subsection{TD-DFT(B)}\label{ss:TDDFTB}

One of the most popular ways to apply time-dependent density functional theory (TD-DFT) in the field of quantum chemistry is without doubt \citeauthor{CasidaTDDFT1995}'s formalism \cite{CasidaTDDFT1995}. Starting from the electron density's linear response to a small perturbation in the external potential, \citeauthor{CasidaTDDFT1995} casts the problem of calculating excitation energies and excited states into an eigenvalue equation in the~$N_\text{trans} = N_\text{occ} N_\text{virt}$ dimensional space of single orbital transitions~$\hat c^\dagger_a c^{\phantom\dagger}_i\ket{\Psi_0}$, where~$\ket{\Psi_0}$ is the Slater determinant of the occupied Kohn-Sham orbitals. The eigenvalue problem can be written as
\begin{equation}\label{eq:CasidasEquation}
\mat \Omega \vec F_I = \Delta_I^2 \vec F_I \; ,
\end{equation}
where~$\Delta_I$ is the excitation energy. The elements of the matrix~$\mat \Omega$ are given by
\begin{equation}
\Omega_{ia,jb} = \delta_{ij} \delta_{ab} \Delta_{ia}^2  + 4 \sqrt{\Delta_{ia}\Delta_{jb}} K_{ia,jb} \; ,
\end{equation}
where we have abbreviated $\Delta_{ia} = \varepsilon_a - \varepsilon_i$. We adopt the usual convention of using the indices~$i,j$ for occupied and~$a,b$ for virtual orbitals. The form of the so-called coupling matrix~$\mat K$ depends on the multiplicity of the excited state. Neglecting spin-orbit coupling, only the singlet excitations are relevant for the calculation of the absorption spectrum. We therefore restrict our discussion to the singlet case, for which the coupling matrix is given by 
\begin{align}
K_{ia,jb} &= \int \dd^3 \vec r \int \dd^3 \pvec r \phi_i(\vec r) \phi_a(\vec r) \\
& \hspace{67.5pt} f_\mathrm{Hxc}[\rho^\mathrm{GS}](\vec r, \pvec r) \; \phi_j(\pvec r) \phi_b(\pvec r) \; . \nonumber
\end{align}
Once the eigenvalue equation~\eqref{eq:CasidasEquation} has been solved, information about the excited state can be extracted from the eigenvectors~$\vec F_i$. Following~\citeauthor{CasidaTDDFT1995}, we use the components of the eigenvector~$\vec F$ to expand the excited state~$\ket{\Psi_I}$ in single orbital excitations relative to the Kohn-Sham Slater determinant~$\ket{\Psi_0}$.
\begin{equation}
\ket{\Psi_I} = \sum_{ia} \sqrt{\frac{2\Delta_{ia}}{\Delta_I}} F_{ia,I} \; \hat c^\dagger_a \hat c^{\phantom\dagger}_i \ket{\Psi_0}
\end{equation}
While the resulting $\ket{\Psi_I}$ should only be viewed as approximation to the true excited state, the transition dipole moment~$\vec d_I$ of the excitation can be calculated as a linear combination of the transition dipole moments~$\vec d_{ia}$ of these single orbital transitions.
\begin{align}
\label{eq:LROsciStrength} \vec d_I = \braket{\Psi_0 | \vec r | \Psi_I} &= \sum_{ia} \sqrt{\frac{2\Delta_{ia}}{\Delta_I}} F_{ia,I} \, \vec d_{ia} \\
\text{with} \qquad \vec d_{ia} &= \braket{\phi_i|\vec r|\phi_a}
\end{align}
The oscillator strength~$f_I$ of the excitation and thereby the absorption spectrum is then easily obtained from
\begin{equation}
f_I = \frac{2}{3} \Delta_I \abs{ \vec d_I }^2 \; .
\end{equation}

While a direct solution of equation~\eqref{eq:CasidasEquation} is in principle possible, the need to store the~$N_\text{trans}^2$ elements of~$\mat \Omega$ in practice limits the size of the treatable systems. In the common case that only~$N_\text{excit} \ll N_\text{trans}$ lowest excitations are needed, this problem can be overcome by the use of iterative eigensolvers, which only need to multiply~$\mat \Omega$ with a set of trial vectors, without ever storing~$\mat \Omega$ explicitly. Not storing the elements of~$\mat \Omega$ implies that they have to be recalculated on-the-fly for every iteration of the eigensolver. The diagonal part of~$\mat \Omega$ is trivial, but the coupling matrix elements involve costly two-center integrals, and even though very efficient methods to calculate these are available~\cite{GisbergenADFTDDFTB1999}, their evaluation still is the major bottleneck in \citeauthor{CasidaTDDFT1995}'s formulation of TD-DFT.  

Time-dependent density functional based tight-binding is a method put forward by \citeauthor{NiehausTDDFTB2001} \cite{NiehausTDDFTB2001,NiehausTDDFTBReview2009,NiehausTDDFTBOnsiteAndFracOcc2013} that builds on SCC-DFTB to approximate the coupling matrix~$\mat K$ to the point where the costly integrals can be parameterized in advance. Let us quickly recapitulate the most important steps of the derivation. First the transition density~$p_{ia}(\vec r) = \phi_i(\vec r) \phi_a(\vec r)$ is decomposed into atomic contributions which are then subjected to a multipole expansion and approximated by their monopolar term.
\begin{equation}\label{eq:TDDFTBAnsatz}
p_{ia}(\vec r) = \sum_\mathcal A p_{ia,\mathcal A}(\vec r) \approx \sum_\mathcal A q_{ia,\mathcal A} \xi_\mathcal A(\vec r)
\end{equation}
Here the~$\xi_\mathcal A(\vec r)$ are the same atom centered functions that are used in the SCC extension of ground state DFTB, and the atomic transition charges~$q_{ia,\mathcal A}$ are calculated from the coefficient and overlap matrices through
\begin{equation}\label{eq:AtomTransCharge}
q_{ia,\mathcal A} = \frac{1}{2} \sum_{\mu \in \mathcal A} \sum_{\nu} \Big( c_{\mu i} S_{\mu \nu} c_{\nu a} + c_{\nu i} S_{\nu \mu} c_{\mu a} \Big) \; .
\end{equation}
Note that the definition of the atomic transition charges also makes it straightforward to calculate the transition dipole moments of the single orbital transitions:
\begin{equation}
\vec d_{ia} = \sum_\mathcal A q_{ia,\mathcal A} \vec R_\mathcal A
\end{equation}
Inserting equation~\eqref{eq:TDDFTBAnsatz} into the expression for the coupling matrix elements yields
\begin{equation}
K_{ia,jb} = \sum_\mathcal{AB} q_{ia,\mathcal A} \tilde{\gamma}_\mathcal{AB} q_{jb,\mathcal B} \; ,
\end{equation}
where the atomic coupling matrix~$\tilde{\mat \gamma}$ is given by
\begin{equation}
\tilde{\gamma}_\mathcal{AB} = \int \hspace{-2.11pt} \dd^3 \vec r \int \hspace{-2.11pt} \dd^3 \pvec r \; \xi_\mathcal A(\vec r) \; f_\mathrm{Hxc}[\rho^\mathrm{GS}](\vec r, \pvec r) \; \xi_\mathcal B(\pvec r) .
\end{equation}
Comparison with equation~\eqref{eq:DFTBGroundStateGamma} reveals that~$\mat \gamma$ and~$\tilde{\mat \gamma}$ only differ in the density at which the derivative of the exchange-correlation energy functional~$E_\mathrm{xc}[\rho]$ is evaluated. At this point \citeauthor{NiehausTDDFTB2001} argue that the second derivative of the exchange-correlation energy is short ranged and therefore only contributes to the on-site elements~$\tilde{\gamma}_\mathcal{AA}$, which are then in analogy to ground state SCC-DFTB approximated by the Hubbard parameters.~\cite{SeifertSCCDFTB1998} \citeauthor{NiehausTDDFTBOnsiteAndFracOcc2013} furthermore show that neglecting the dependence of the Hubbard parameters on the atomic charges is consistent within a linear response treatment based on ground state SCC-DFTB~\cite{NiehausTDDFTBOnsiteAndFracOcc2013}. Using the Hubbard parameters of the neutral atoms reduces the atomic coupling matrix $\tilde{\mat \gamma}$ to the $\mat \gamma$~matrix from ground state SCC-DFTB, which then leads to a simple equation for the matrix~$\mat \Omega$.
\begin{equation}\label{eq:OmegaMatrixDFTB}
\Omega_{ia,jb} =  \delta_{ij} \delta_{ab} \Delta_{ia}^2  + 4 \sqrt{\Delta_{ia}\Delta_{jb}} \sum_\mathcal{AB} q_{ia,\mathcal A} \gamma_\mathcal{AB} q_{jb,\mathcal B}
\end{equation}
Note that the orbital energy differences~$\Delta_{ia}$ as well as the coefficient matrix~$\mat C$ and the overlap matrix~$\mat S$ can easily be extracted from any DFTB ground state calculation, and that no TD-DFTB specific parameters are needed since the $\mat{\gamma}$~matrix already had to be parameterized within the SCC-DFTB method. TD-DFTB can therefore immediately be applied to any system for which ground state SCC-DFTB parameters are available. It would be beyond the scope of this article to validate the TD-DFTB method itself. Such studies have of course been performed~\cite{NiehausTDDFTB2001,NiehausTDDFTBReview2009,GaussianTDDFTB2011,NiehausTDDFTBOnsiteAndFracOcc2013} and while the approximations made in TD-DFTB seem drastic at first sight, the overall accuracy of the method has been found to be promising and TD-DFTB has since seen a wide variety of applications.~\cite{tddftbapp1_doi:10.1021/jp026752s,tddftbapp2_PhysRevB.73.205312, tddftbapp3_doi:10.1021/jp071125u,tddftbapp4_doi:10.1021/ct700041v,tddftbapp5_doi:10.1021/jp065704v,tddftbapp6_10.1063/1.2715101,tddftbapp7_doi:10.1063/1.2940735,BonacicNonAdMDWithTDDFTB2009,tddftbapp9_PSSB:PSSB201100719,tddftbapp10_Fan201417}

In summary, TD-DFTB is a computationally rather simple approximation to TD-DFT where the computational bottleneck is the size of the response matrix~$\mat \Omega$ and the calculation of its eigenvectors. In the next section we will present computational methods to solve the TD-DFTB equations efficiently.

\section{Computational methods}\label{s:ComputationalMethods}

As both the number of occupied~$N_\text{occ}$ and the number of virtual orbitals~$N_\text{virt}$ grow linearly with the number of atoms~$N_\text{atom}$, the total number of single orbital transitions~$N_\text{trans} = N_\text{occ} N_\text{virt}$ increases quadratically with the system size. This in practice limits the size of the systems treatable with TD-DFT(B), which uses the single orbital transitions as the basis of the space in which \citeauthor{CasidaTDDFT1995}'s eigenvalue equation~\eqref{eq:CasidasEquation} has to be solved. An exact diagonalization of the full matrix~$\mat \Omega$ is only possible for the smallest systems, as the memory required to store~$\mat \Omega$ scales as~$\mathcal O(N_\text{trans}^2)$, which equates to a prohibitive~$\mathcal O(N_\text{atom}^4)$ scaling. A lot of applications only need a small part of the spectrum at its low energy end, so it is possible to use iterative eigensolvers that avoid storage of the full matrix~$\mat \Omega$ in favor of a series of matrix-vector multiplications. Especially popular in the context of TD-DFT(B) is a class of methods based on an idea by Davidson~\cite{Davidson1975}, in which the eigenvalue problem is solved approximately in a small subspace, which is then iteratively extended and refined to include the desired eigenvectors within a certain accuracy. There is a multitude of different Davidson based diagonalization algorithms and reviewing them would be beyond the scope of this article. As the eigensolver for TD-DFTB calculations we use a variant of the GD+$k$ method developed by Stathopoulos and Saad~\cite{StathopoulosGD+k1998} and implemented in the PRIMME library~\cite{StathopoulosPRIMME2010}. While the eigensolver internally needs to store the subspace basis, this required memory scales as~$\mathcal O(N_\text{trans})$ and is often negligible in comparison to the $(N_\text{trans} \times N_\text{excit})$~matrix of the desired eigenvectors.

\subsection{Efficient implementation of the matrix-vector multiplication}\label{ss:MatVecMulImpl}

Eigensolvers based on the Davidson method~\cite{Davidson1975} solve the eigenvalue problem approximately in a small subspace which is then iteratively expanded by adding new basis vectors until it contains the desired eigenvectors. They only use the matrix they diagonalize in terms of a matrix-vector multiplication with the newly added basis vectors. In practice this is actually a matrix-matrix multiplication as it is common to add~$N_\text{block} \geq 1$ basis vector per iteration. This is known as the block Davidson method which was proposed by Liu~\cite{LiuBlockDavidson1978} as a method to increase computational efficiency and to improve convergence for degenerate eigenvalues. In case of the block Davidson method the only part of the algorithm that is referencing the original matrix~$\mat \Omega$ can be written as
\begin{equation}\label{eq:MatVecMult}
\mat R = \mat \Omega \mat T \; ,
\end{equation}
where~$\mat T$ is an $(N_\text{trans} \times N_\text{block})$~matrix whose columns are the newly added basis vectors. We want to discuss the implementation of this matrix-vector multiplication in some more detail now, as it is crucial to the performance of the entire TD-DFTB method.

As storage of the full matrix~$\mat \Omega$ is certainly impossible -- hence the iterative solution in the first place -- we need to recalculate its elements during every matrix-vector multiplication. We can, however, precalculate a set of smaller auxiliary objects from which $\mat \Omega$ can be obtained more quickly.

Inserting equation~\eqref{eq:OmegaMatrixDFTB} into~\eqref{eq:MatVecMult} it is easy to see that one can precalculate a scaled version of the atomic transition charges~$q_{ij,\mathcal A}$ in order to turn the multiplication with the large coupling matrix~$\mat K$ into a series of matrix-matrix multiplications involving only smaller matrices.
\begin{align}
\label{eq:precalc_writtenout} R_{ia,I} &= \Delta_{ia}^2 T_{ia,I} +  4 \sum_\mathcal A \underbrace{\sqrt{\Delta_{ia}} \, q_{ia,\mathcal A}}_{h_{ia,\mathcal A}} \\[-4pt]
& \hspace{86pt} \sum_{\mathcal B} \gamma_\mathcal{AB} \sum_{jb} \underbrace{\sqrt{\Delta_{jb}} \, q_{jb,\mathcal B}}_{h_{jb,\mathcal B}} T_{jb,I} \nonumber \displaybreak[0]\\
\label{eq:precalc} \mat R &= \mathrm{diag}\left(\Delta_{ia}^2\right) \mat T + 4 \mat h \mat \gamma \mat h^T \mat T
\end{align}
Here~$\mat h$ is of size~$(N_\text{trans} \times N_\text{atom})$ whereas $\mat \gamma$ is~$(N_\text{atom} \times N_\text{atom})$. In order to ensure the overall cubic scaling of the matrix-matrix products we need to evaluate the subexpressions via temporary objects.
\begin{align}
\label{eq:matvec_temp1} X_{\mathcal B I} &= \sum_{jb} \displaystyle h_{jb,\mathcal B} T_{jb,I} \displaybreak[0]\\
Y_{\mathcal A I} &= \sum_\mathcal B \gamma_\mathcal{AB} X_{\mathcal B I} \displaybreak[0]\\
\label{eq:matvec_temp3} R_{ia,I} &= \Delta_{ia}^2 T_{ia,I} + 4 \sum_\mathcal A h_{ia,\mathcal A} Y_{\mathcal A I}
\end{align}
Here the first and third step scale as $\mathcal O(N_\mathrm{trans} N_\mathrm{atom} N_\mathrm{block})$, whereas $\mathcal O(N_\mathrm{atom}^2 N_\mathrm{block})$ operations are needed for the intermediate step, which is negligible since~$N_\mathrm{atom} \ll N_\mathrm{trans}$. Note that this is only the scaling of a single matrix-vector product, which is different from the total time spent in matrix-vector products: Considering the entire calculation instead of the single product, the total number of trial vectors required for convergence is roughly linear in the number of requested excitations~$N_\mathrm{excit}$, no matter how the trial vectors are blocked during the multiplications. Ergo, it is more insightful to consider the scaling of the total time spent performing matrix-vector products, which is~$\mathcal O(N_\mathrm{trans} N_\mathrm{atom} N_\mathrm{excit})$. The scaling behavior of the different operations involved in TD-DFTB is summarized in table~\ref{tab:scaling_time}.

Equation~\ref{eq:precalc} provides an extremely fast way to perform the matrix-vector product as only basic linear algebra operations are used which can be offloaded to highly optimized libraries. If for large systems the matrix~$\mat h$ of the scaled atomic transition charges becomes too large to be stored though, it is necessary to recalculate its elements during the matrix-vector multiplications. Looking again at equation~\eqref{eq:AtomTransCharge} it is easy to see that the sum over~$\nu$ is just a regular matrix-matrix multiplication between the overlap matrix~$\mat S$ and the coefficient matrix~$\mat c$.
\begin{align}
h_{ia,\mathcal A} &= \frac{1}{2} \sqrt{\Delta_{ia}} \sum_{\mu \in \mathcal A} \sum_{\nu} \Big( c_{\mu i} S_{\mu \nu} c_{\nu a} + c_{\nu i} S_{\nu \mu} c_{\mu a} \Big) \nonumber \\
&= \frac{1}{2} \sqrt{\Delta_{ia}} \sum_{\mu \in \mathcal A} \Big( c_{\mu i} \Theta_{\mu a} + c_{\mu a} \Theta_{\mu i} \Big)
\end{align}
The product matrix~$\mat \Theta = \mat S \mat c$ can be calculated in advance and stored instead of~$\mat S$ without additional memory in a full matrix storage implementation. The calculation of the scaled atomic transition charge~$h_{ij,\mathcal A}$ then only contains a sum over the basis functions centered on atom~$\mathcal A$, which is usually a small number due to the minimal basis set and the large frozen core typically used in DFTB calculations. Note that precalculating~$\mat \Theta$ makes it possible to calculate the elements of~$\mat h$ in a system-independent constant time, so that evaluating them on-the-fly does not change the scaling of the matrix-vector multiplication but only increases the prefactor.

In case of precalculated atomic transition charges one can rely on standard libraries to perform the parallelization of the matrix-vector product. This is no longer true for on-the-fly calculated transition charges, where one has to parallelize equation~\eqref{eq:matvec_temp1} and~\eqref{eq:matvec_temp3} manually. Both equations can easily be parallelized, but one has to pay attention to distribute the work such that each scaled atomic transition charge~$h_{ij,\mathcal A}$ is in total only calculated once per step: The element~$X_{\mathcal B I}$ in equation~\eqref{eq:matvec_temp1} depends both on the atom~$\mathcal B$ as well as the trial vector index~$I$, but the element~$h_{jb,\mathcal B}$ only depends on the atom~$\mathcal B$. Therefore, the parallelization is chosen to be done over the atoms~$\mathcal B$ since parallelizing over the index~$I$ would require every processor to calculate~$h_{jb,\mathcal B}$. The matrix-matrix product in equation~\eqref{eq:matvec_temp3} is chosen to be parallelized via the transition index~$ia$ for the exact same reason. In summary, recalculating the atomic transition charges on-the-fly during the matrix-vector multiplications removes the need to store the matrix~$\mat h$ of size~$(N_\text{trans} \times N_\text{atom})$. The storage required for the coefficient matrix~$\mat c$ and product matrix~$\mat \Theta$ can usually be neglected compared to the $(N_\text{trans} \times N_\text{excit})$~matrix of the desired eigenvectors. The memory requirements for all the different methods are summarized in table~\ref{tab:scaling_memory}.

At this point it is necessary to mention that while its performance is certainly important, the matrix-vector multiplication is not always the bottleneck of the Davidson eigensolver. The reason for this is that in order to find the $N$th~eigenvector it is necessary to orthonormalize it against the~$N-1$ already known eigenvectors. This has an~$\mathcal O(N_\mathrm{trans} N_\mathrm{excit}^2)$ scaling which for large~$N_\mathrm{excit}$ dominates over the $\mathcal O(N_\mathrm{trans} N_\mathrm{atom} N_\mathrm{excit})$ scaling of the matrix-vector multiplication.

\subsection{Basis size reduction by transition selection}\label{ss:TransitionSelection}

While iterative eigensolvers make TD-DFTB calculations of larger molecules possible in the first place, the huge dimension~$N_\text{trans}$ of the single orbital transition space still limits the size of the treatable systems. It is therefore worthwhile to investigate the possibility of working in a subspace of single orbital transitions in which the (approximately) same result can be obtained using fewer transitions.

The most obvious way to reduce the basis size is a truncation in energy: As the iterative solution of the eigenvalue problem only targets a few of the lowest eigenvectors of a typically diagonally dominant matrix, the eigenvector can be expected to have little overlap with basis vectors for which the diagonal element is large. In physical terms this just means that the transitions from the lowest most tightly bound molecular orbitals to the highest virtuals will usually not contribute to the lowest excitations, which mostly consist of transitions close to the HOMO-LUMO gap.

Our target application of TD-DFTB are UV/Vis absorption spectra, for which the solution of \citeauthor{CasidaTDDFT1995}'s eigenvalue equation~\eqref{eq:CasidasEquation} produces the excitation energies~$\Delta_I$, while the corresponding oscillator strengths~$f_I$ can be calculated through equation~\eqref{eq:LROsciStrength}. Together these can immediately be used to plot a stick-like spectrum, that using Dirac's $\delta$-distribution could be written as
\begin{equation}
A_\mathrm{stick}(E) = \sum_I f_I \; \delta(E - \Delta_I) \; .
\end{equation}
As these spectra are both hard to interpret and unrealistic, it is common practice to artificially introduce line broadening through a convolution with a peaked function~$\Gamma(E)$.
\begin{equation}
\begin{split}
A_\mathrm{broad.}(E) &= \int \dd E' \; \Gamma(E' - E)  A_\mathrm{stick}(E') \\
&= \sum_I f_I \; \Gamma(E - \Delta_I)
\end{split}
\end{equation}
Both Gaussian and Lorentzian functions are common choices for~$\Gamma(E)$. As the absorption peaks are scaled with the oscillator strength~$f_I$ of the excitation, the absorption spectrum is mostly determined by the excitations which have a large oscillator strength. Looking at equation~\eqref{eq:LROsciStrength} for the transition dipole moment of the excitations, it is easy to see that single orbital transitions with a small transition dipole moment~$\vec d_{ia}$ contribute little to the transition dipole moment of the excitation~$\vec d_I$, and hence its oscillator strength~$f_I$. Consequently it appears to be a reasonable approximation to remove those single orbital transitions from the basis for which the oscillator strength~$f_{ia}$ is small. Note that this is an approximation, as even leaving out a single orbital transition with~$f_{ia} = 0$ might still influence the oscillator strength~$f_I$ through an overall change in the corresponding eigenvector~$\vec F_I$. The benefit of removing single orbital transitions with small oscillator strengths~$f_{ia}$ goes beyond the obvious reduction in computational effort associated with the smaller dimension of the eigenvalue problem: As one is essentially working in the oscillator strength carrying subspace, many of the excitations with small oscillator strength~$f_I$ are also removed from the final spectrum, making it possible to calculate the absorption spectrum in a fixed energy window with fewer excitations. It is in fact an all too common problem that a large number of excitations has to be calculated in order to cover the energy window of interest, while only a few of them actually determine the shape of the absorption spectrum due to their large oscillator strength~$f_I$.

For a direct diagonalization of the $\mat \Omega$~matrix it is obvious that the relative reduction in basis size translates quadratically into memory savings and cubically into reduced processor time, compare table~\ref{tab:scaling_time} and~\ref{tab:scaling_memory}. For the iterative solvers the situation is more complicated due to the fact that the number of excitations that have to be calculated within a fixed energy interval is also reduced: Depending on whether the matrix-vector multiplication or the orthonormalization of the subspace basis is the bottleneck, the relative reduction in basis size will translate either quadratically or cubically into reduced processor time.

The idea to reduce the number of considered single orbital transitions is not entirely new: A truncation of the single orbital transition space based on orbital localization has successfully been used by \citeauthor{BesleyTruncationOnSurfaces2004} for the special cases of molecules in solution and on surfaces \cite{BesleyTruncationOnSurfaces2004}. The more generally applicable truncation in energy or oscillator strength has 
recently been also proposed and tested in the PhD thesis of~\citeauthor{DominguezPhDThesis2014}, but no in-depth evaluation of the method was performed \cite{DominguezPhDThesis2014}. In the next section we will assess the validity of the approximations introduced by truncating the basis in energy or oscillator strength, and we will show that these techniques can at negligible loss in accuracy lead to orders of magnitude reductions in computer time and required memory.

\section{Examples}\label{s:Examples}

The accuracy loss due to the additional approximation introduced by the truncation of the single orbital transition basis certainly needs to be investigated in order to judge whether these approximations can be used in practice. Furthermore we need to determine to which extent the loss in accuracy is justified by the computational benefits of truncation. Detailed timings of the various example calculations can be found in table~\ref{tab:timings}. Note that we can use arbitrary units as we are only comparing theoretical data in these examples, for comparison with experimental data one may insert the appropriate prefactors for the desired unit system. 

\subsection{Fullerene C$_{60}$}

The fullerene~C$_{60}$ was used by \citeauthor{NiehausTDDFTB2001} in the original TDDFTB article \cite{NiehausTDDFTB2001} as a benchmark to judge the quality of the approximations introduced by TD-DFTB in general. The authors found that the inclusion of coupling between the single orbital transitions is crucial in the description of the optical properties of~C$_{60}$ and that TD-DFTB qualitatively reproduces the main features of the experimental spectrum~\cite{C60Exp}.

We have performed a series of calculations with differently truncated single orbital transition spaces. With 4 valence electrons per atom, the~C$_{60}$ molecule has 120 occupied and 120 virtual orbitals (assuming a minimal basis), which results in a total of 14400 single orbital transitions. For this rather small number of transitions it is still possible to perform an exact diagonalization of the $\mat \Omega$~matrix. We used the carbon parameters included in the mio-1-1 parameter set.~\cite{SeifertSCCDFTB1998}

\begin{figure}[tb]
\includegraphics[width=\columnwidth]{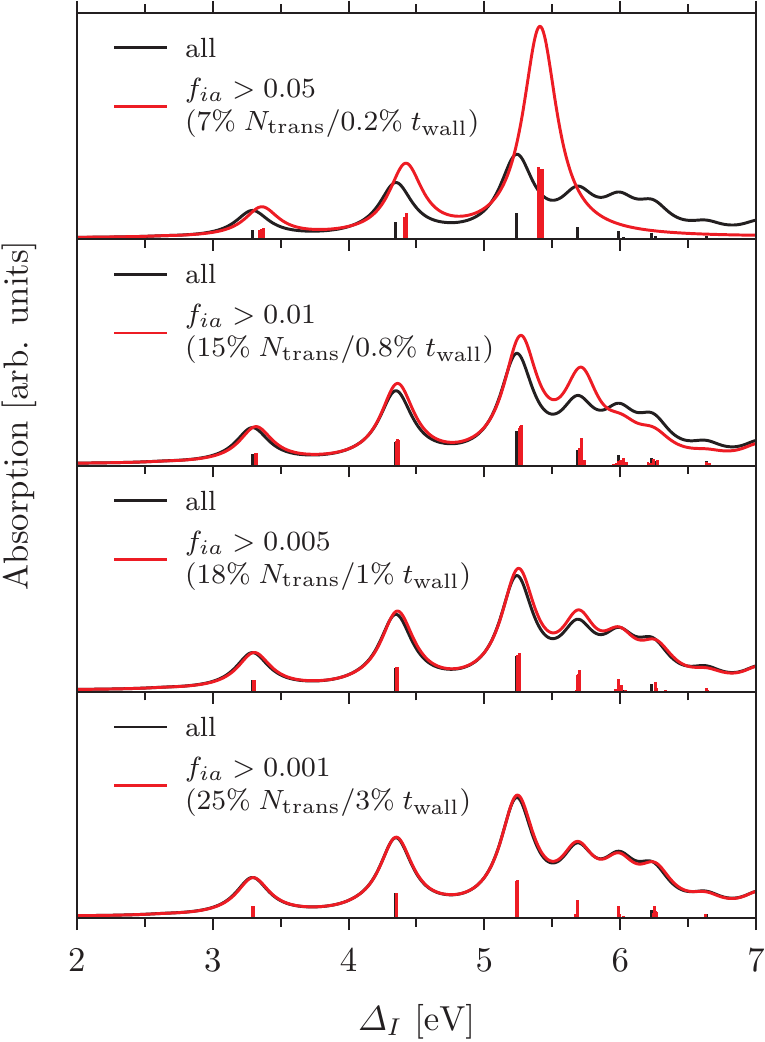}
\caption{\label{fig:C60_iseffect}TD-DFTB calculated absorption spectra of C$_{60}$ fullerene with different intensity selection thresholds. The percentage in the parentheses is the size of the remaining basis and the required computational time relative to the full calculation.}
\end{figure}

Figure~\ref{fig:C60_iseffect}
shows absorption spectra calculated using a basis from which single orbital transitions with an oscillator strength~$f_{ia}$ smaller than a user defined threshold~$f_{ia}^\mathrm{min}$ have been removed. As expected the quality of the approximation decreases as the threshold~$f_{ia}^\mathrm{min}$ is increased and more and more of the single orbital transitions are removed. Note that there is a slight blueshift of the main peaks for larger~$f_{ia}^\mathrm{min}$. Looking at the bottom plot in figure~\ref{fig:C60_iseffect} one can see that a large part of the basis does not seem to contribute to the absorption spectrum at all, as a threshold of~$f_{ia}^\mathrm{min} = 0.001$ already removes three quarters of all single orbital transitions while leaving the obtained absorption spectrum practically unchanged. The reason for this is that for the highly symmetric fullerene~C$_{60}$ there are a lot of single orbital transitions where the transition dipole moment~$\vec d_{ia}$ and hence the oscillator strength~$f_{ia}$ is zero purely due to symmetry. This is a great advantage for the use of intensity selection and leads to a wall time reduction by two orders of magnitude at a negligible loss in accuracy for a selection threshold of~$f_{ia}^\mathrm{min} = 0.002$. We will later look at less symmetric examples though, where this does not play a role.

\begin{figure}[tb]
\includegraphics[width=\columnwidth]{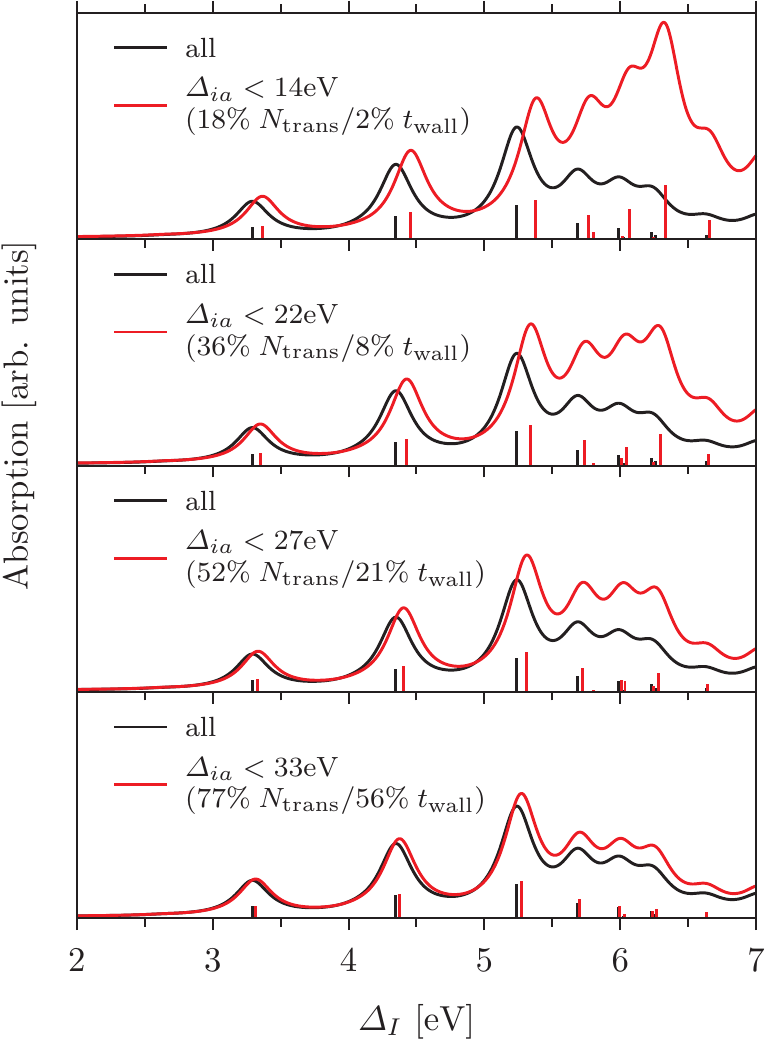}
\caption{\label{fig:C60_eteffect}TD-DFTB calculated absorption spectra of C$_{60}$ fullerene with different energy truncation thresholds. The percentage in the parentheses is the size of the remaining basis and the required computational time relative to the full calculation.}
\end{figure}

Figure~\ref{fig:C60_eteffect}
shows absorption spectra calculated using a basis from which single orbital transitions with a large orbital energy difference~$\Delta_{ia}$ have been removed. It is evident that truncation of the basis in energy has a relatively large effect on the absorption spectrum, at least compared to the intensity selection. While the number of peaks is preserved upon energy truncation, they are subject to a sizeable blueshift and their relative oscillator strength is not well preserved. Overall this results in a too strong absorption band around~6eV that does not exist in this form in calculations using the full basis. A possible reason for the mediocre performance of the energy truncation could be the fact that the orbital energy difference directly enters into equation~\eqref{eq:LROsciStrength} for the transition dipole moment~$\vec d_I$ of the linear response excitations, giving high energy transitions a disproportionately large effect on the low energy end of the absorption spectrum, even though the associated eigenvector elements~$F_{ia,I}$ might be rather small. A major disadvantage of the truncation in energy compared to the intensity selection is that it does not reduce the number of excitations per energy interval, so that for the iterative solver the relative reduction in basis size translates only linearly into memory savings and reduced processor time. Our overall experience is that the truncation in energy introduces non-negligible errors while offering only moderate computational advantages. While it is easily possible to combine truncation in energy with truncation in oscillator strength, we have found that even this is consistently outperformed by pure intensity selection on which we will therefore focus in the remainder of this article.

\subsection{Ir(ppy)$_3$}

The compound Tris(2-phenylpyridine)iridium, abbreviated as Ir(ppy)$_3$, has recently been discussed in the context of highly efficient organic light emitting diodes~\cite{irppy3_0}. There are two geometrical isomers, facial~(\textit{fac}-Ir(ppy)$_3$) and meridional~(\textit{mer}-Ir(ppy)$_3$), where the former is lower in energy. We will therefore only discuss the \textit{fac}-Ir(ppy)$_3$~isomer. While the triplet excitations of Ir(ppy)$_3$ are technically more interesting due to their role in the process called triplet-harvesting~\cite{tripletharvesting}, theoretical as well as experimentally obtained absorption spectra can also be found in the literature~\cite{irppy3_1,irppy3_2}. These show two absorption bands around 3.5eV and 5eV. The former band has been found to originate from metal to ligand charge transfer, while the latter more intense band around 5eV has been attributed to $\pi-\pi^*$~excitations in the ligand.

We performed TD-DFTB calculations on \textit{fac}-Ir(ppy)$_3$ using the parameters developed by \citet{WahiduzzamanQuasinano2013}, which include parameters for the central Iridium atom. Ir(ppy)$_3$ has a total of 7830 single orbital transitions so that the $\mat \Omega$~matrix can easily be diagonalized exactly.

\begin{figure}[tb]
\includegraphics[width=\columnwidth]{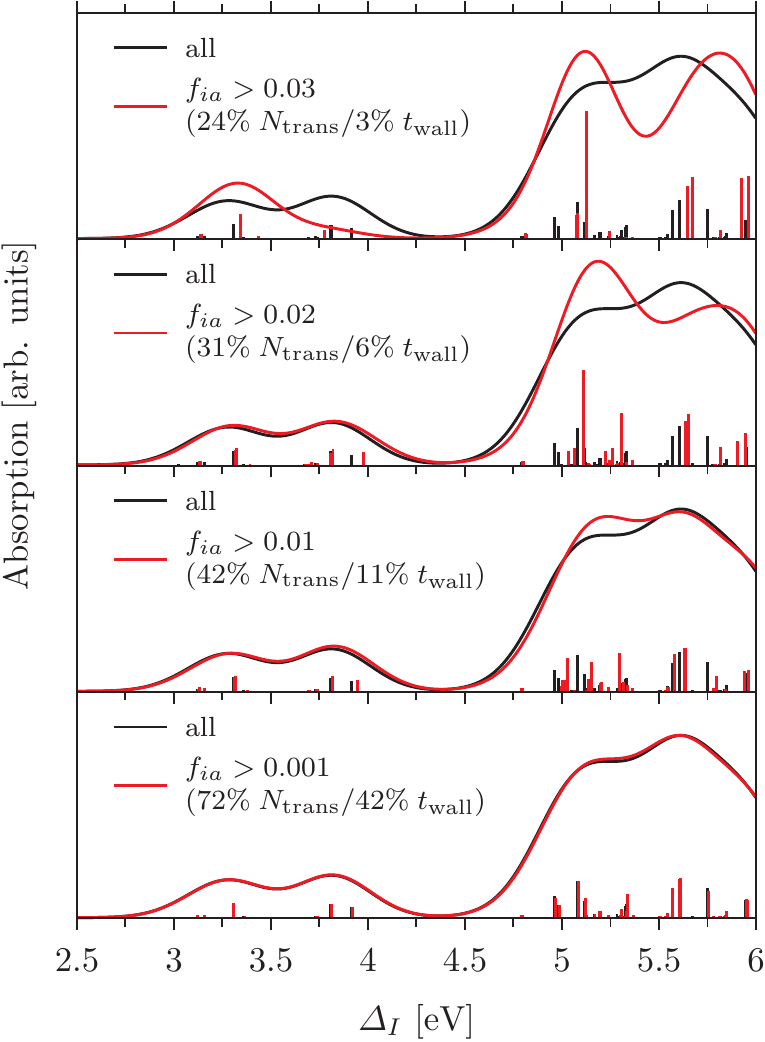}
\caption{\label{fig:irppy3_iseffect}TD-DFTB calculated absorption spectra of \textit{fac}-Ir(ppy)$_3$ with different intensity selection thresholds. The percentage in the parentheses is the size of the remaining basis and the required computational time relative to the full calculation.}
\end{figure}

Figure~\ref{fig:irppy3_iseffect}
shows the TD-DFTB calculated absorption spectrum obtained with intensity selection at different oscillator strength thresholds. TD-DFTB reproduces the general shape of the TD-DFT calculated absorption spectra published by \citet{irppy3_1}, though the more intense band at higher energies is blueshifted by about 0.5eV. As was the case for the fullerene example, the absorption spectrum is practically unchanged when imposing an intensity selection threshold of~$f_{ia}^\mathrm{min} = 0.001$. In contrast to the fullerene example though, the resulting reduction of the basis size is far less drastic: A threshold of~$f_{ia}^\mathrm{min} = 0.001$ removes 75\% of the fullerene single orbital transitions, but only 28\% of the transitions in Ir(ppy)$_3$. This is due to the fact that the less symmetric \textit{fac}-Ir(ppy)$_3$ does not have any single orbital transitions whose transition dipole moment vanishes purely due to symmetry. Increasing the selection threshold decreases the quality of the approximation as seen in figure~\ref{fig:irppy3_iseffect}, but it is not until~$f_{ia}^\mathrm{min} = 0.03$ (which results in a 76\% reduction) that the spectrum starts to become qualitatively different. Overall, carefully used intensity selection in case of~\textit{fac}-Ir(ppy)$_3$ provides sizable computational advantages with wall time reductions up to one order of magnitude and little loss of accuracy.

\subsection{Ubiquitin}

Ubiquitin~\cite{ubq_discovery} is an extremely common small protein that has various regulatory functions in almost all eukaryotic cells.~\cite{physrev.00027.2001, Schnell19092003, Mukhopadhyay12012007} It has recently been used as an example system for UV/VIS spectroscopy of entire proteins in gas phase~\cite{ubq} so that both experimentally observed as well as theoretically calculated absorption spectra are available.~\cite{ubq_work} The low energy part of the ubiquitin absorption spectrum is dominated by absorption in the single tyrosine amino acid, so that \citeauthor{ubq_work} were able to calculate ubiquitin's absorption spectrum using a QM/MM approach~\cite{ubq_work}, where the tyrosine chromophore is embedded into a classical environment (modeled with the Amber force field~\cite{CornellAMBER1995}), while the chromophore itself is treated quantum mechanically with TD-DFT (B3LYP/aug-cc-pvdz). 

\begin{figure}[tb]
\includegraphics[width=\columnwidth]{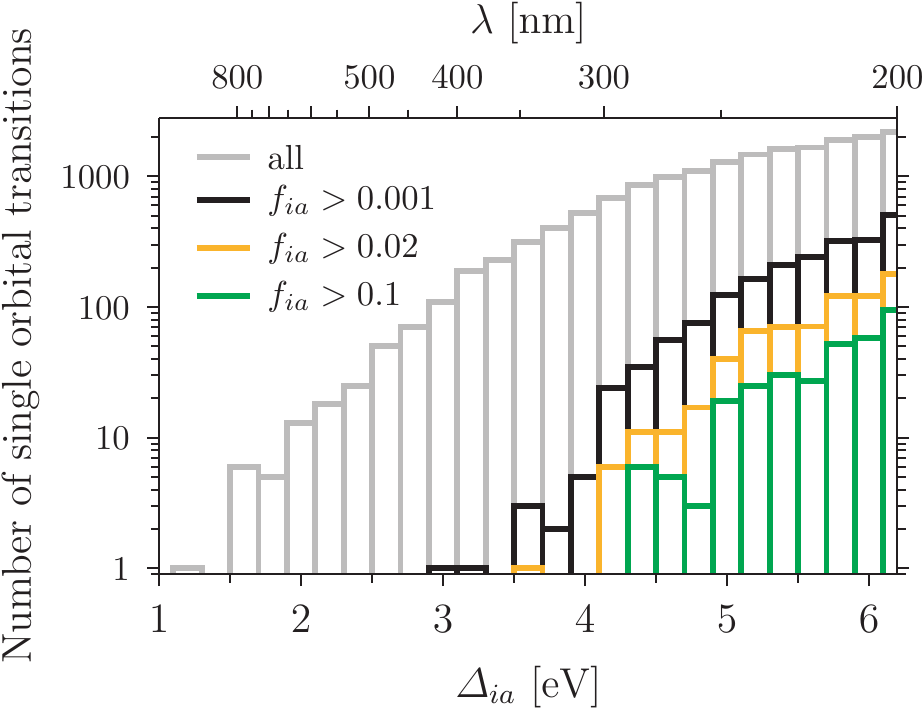}
\caption{\label{fig:ubq_sohist}Number of single orbital transitions per energy interval for ubiquitin for different intensity selection thresholds. Note the large number of low intensity transitions below 3.5eV that is removed by even a small threshold.}
\end{figure}

We performed TD-DFTB calculations using the mio-1-1 parameter set~\cite{SeifertSCCDFTB1998} based on the PBE functional~\cite{PerdewBurkeErnzerhofPBEXCFunc1996}. For such a large system the iterative solution of the eigenvalue problem is essential, but with 1231 atoms and in total $2\,284\,880$~single orbital transitions the 22~gigabyte matrix of atomic transition charges can still be precalculated and stored in memory, so that the matrix-vector multiplication can be implemented as equation~\ref{eq:precalc}. If one attempts to calculate the absorption spectrum up to~200nm without using intensity selection, one quickly finds that there are almost $16\,000$ single orbital transitions within this window, so that an equally large number of excitations would have to be calculated to get the interesting part of the absorption spectrum. With 18~megabyte of memory per eigenvector, this would require almost 290~gigabyte to store the solution, which is rather excessive. Analysis of the single orbital transitions reveals though, that many of them have a very small oscillator strength. This is visualized in figure~\ref{fig:ubq_sohist}
where the number of single orbital transitions per energy interval is plotted for different oscillator strength thresholds. It is evident that almost all single orbital transitions below~4eV have an oscillator strength~$f_{ia} < 0.001$ and would be removed if intensity selection was applied. Setting a threshold of~$f_{ia}^\mathrm{min} = 0.001$ in total removes~29\% of the single orbital transitions, but looking only at the relevant part of the spectrum up to~200nm it reduces the number of transitions to about~1600, which is a reduction by one order of magnitude. This not only makes the solution much faster, but also only requires memory for 1600~eigenvectors of 13~megabyte each, which is 21~gigabyte in total and certainly manageable.

\begin{figure}[tb]
\includegraphics[width=\columnwidth]{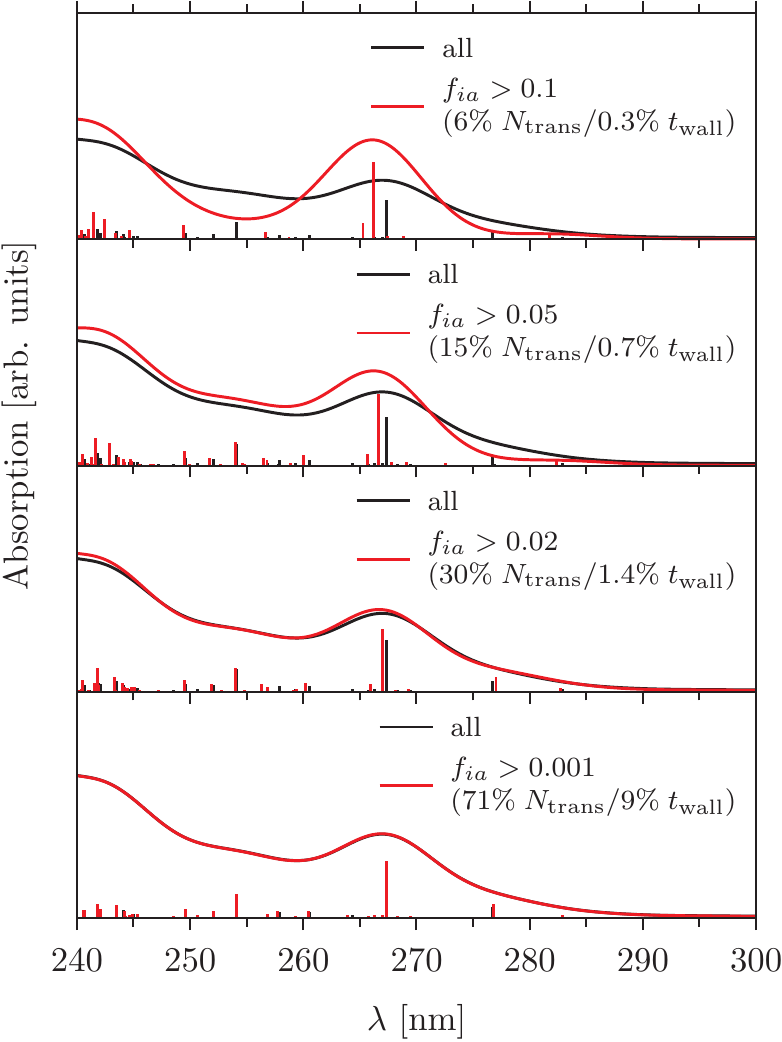}
\caption{\label{fig:ubq_iseffect}TD-DFTB calculated absorption spectra of ubiquitin with different intensity selection thresholds. The percentage in the parentheses is the size of the remaining basis and the required computational time relative to the full calculation. Note that the intensity-selected calculations were run on fewer cluster nodes than the full calculation so that the shown wall times underestimate the speedup. Detailed timings can be found in table~\ref{tab:timings}.}
\end{figure}

The absorption spectrum of ubiquitin calculated using TD-DFTB with different intensity selection thresholds is shown in figure~\ref{fig:ubq_iseffect}.
Except for a slight redshift of the first absorption band around~267nm, TD-DFTB overall very well reproduces the spectrum obtained by \citeauthor{ubq_work}. Concerning the intensity selection, it is especially remarkable that imposing the aforementioned oscillator strength threshold of~$f_{ia}^\mathrm{min} = 0.001$ does not change the resulting absorption spectrum at all, even though it reduces the number of excitations in the shown energy window by more than one order of magnitude. Increasing the threshold to~$f_{ia}^\mathrm{min} = 0.02$ removes~$70\%$ of the basis while still producing an essentially perfect absorption spectrum at a drastically reduced computational cost: While the calculation using the full basis took more than 12 hours and had to be run on 8 cluster nodes due to its substantial memory requirements, the intensity-selected calculation with a $f_{ia}^\mathrm{min} = 0.02$~threshold finished in less than 15 minutes on only two cluster nodes. Further increasing the threshold to~$f_{ia}^\mathrm{min} = 0.05$ the intensity selection's influence on the spectrum becomes more noticable: We observe a slight blueshift and an increase in intensity of the band around~267nm, and for large thresholds we also see some excitations vanish, most notably two relatively intense excitations at~254nm and~277nm, whose disappearance further contributes to making the central absorption band stand out.

The reason why there are so many excitations with practically zero oscillator strength at low energies is that these are mostly charge-transfer excitations, where an electron is transferred from one part of the molecule (the donor) to another part (the acceptor), possibly over a relatively long distance. It is widely known that Kohn-Sham DFT based calculations can drastically underestimate the excitation energies of such charge-transfer excitations, due to the fact that the LUMO energy of the acceptor does not correspond to its electron affinity, as would be correct in case of a charge-transfer excitation where the acceptor essentially gains an additional electron.~\cite{GritsenkoCTFailure2004} It is interesting to note though that charge-transfer excited states typically have a small overlap with the ground state and thereby according to equation~\eqref{eq:LROsciStrength} also a rather small transition dipole moment~\cite{MagyarCtOsciStr2007}. While intensity selection by no means solves the underlying problem of too small charge-transfer excitation energies in Kohn-Sham DFT, it at least helps to alleviate the worst of the associated computational problems for the specific application of calculating electronic absorption spectra.

\begin{figure}[tb]
\vspace{-16.5pt}\includegraphics[width=\columnwidth]{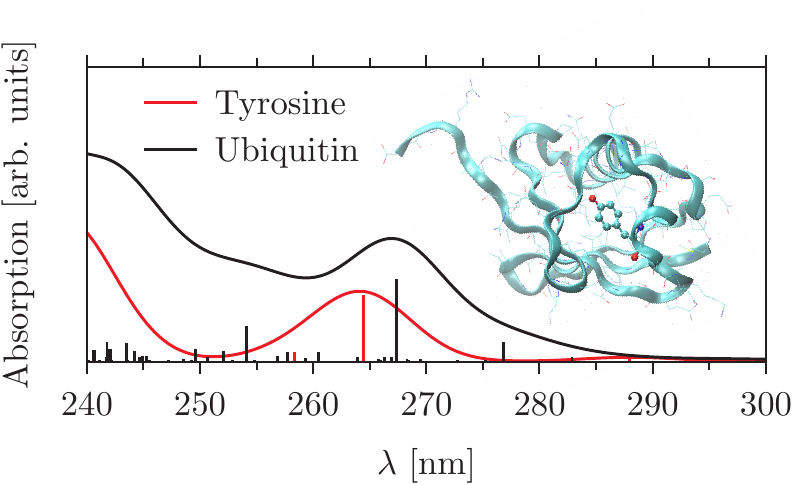}
\caption{\label{fig:ubq_embedding}Comparison of the absorption spectra of tyrosine and ubiquitin. The embedding of the tyrosine into the protein environment produces a slight redshift of the absorption band at~264nm.}
\end{figure}

As a last practical example we have tried to reproduce the spectral shift associated with the inclusion of the tyrosine chromophore into the protein environment. Figure~\ref{fig:ubq_embedding} 
shows the TD-DFTB calculated absorption spectra of ubiquitin and the isolated tyrosine in comparison. TD-DFTB predicts a blueshift of about 4nm upon embedding of the chromophore into the protein environment, which is in agreement with the shift calculated by \citeauthor{ubq_work}. This goes to show that intensity-selected TD-DFTB is a viable alternative to QM/MM methods for the calculation of electronic absorption spectra of large compounds. In addition to the more accurate treatment of the environment, a general advantage of TD-DFTB over QM/MM is that it is much easier to use, as the user does not have to first identify the chromophore and does not have to make decisions on which part to treat quantum mechanically and how to embed it into the classically treated region. 

\subsection{Parallel scaling}

\begin{figure}[tb]
\includegraphics[width=\columnwidth]{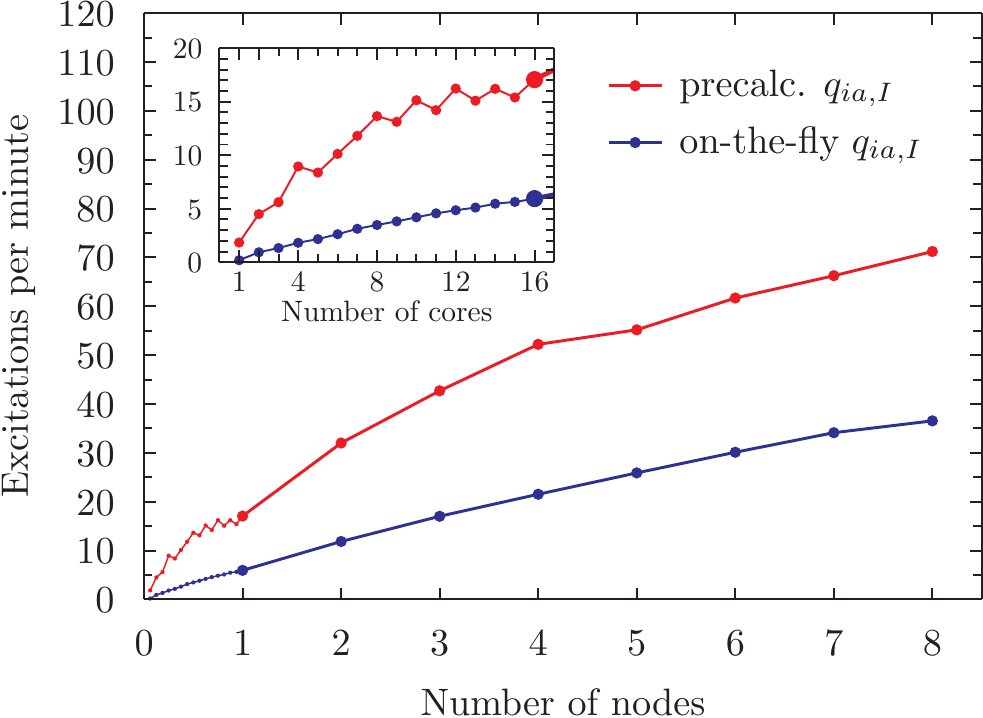}
\caption{\label{fig:ubq_scaling}Parallel scaling of our TD-DFTB implementation for the test case of ubiquitin with an intensity selection threshold of~$f_{ia}^\mathrm{min} = 0.005$. The calculations were performed on 1 to 8 cluster nodes with two octa-core Intel Xeon E5-2650~v2 processors each and 64GB of memory per node.}
\end{figure}

In order to evaluate the performance of our parallel implementation, we have conducted a scaling test for the example calculation of the ubiquitin absorption spectrum with an intensity selection threshold of~$f_{ia}^\mathrm{min} = 0.005$. The scaling test was performed on 1 to 8 cluster nodes with two octa-core Intel Xeon E5-2650~v2 processors each and 64GB of memory per node. The particular threshold of~$f_{ia}^\mathrm{min} = 0.005$ was chosen as it results in a calculation that (using precalculated atomic transition charges) barely fits into the memory of a single node. In this way we conducted the scaling test on the largest system we were able to solve in serial, allowing us to plot the scaling behavior for the entire range from~1 to~128 cores. The result of the scaling test for both precalculated and on-the-fly atomic transition charges is shown in figure~\ref{fig:ubq_scaling}.

For precalculated transition charges we observe a good scaling both within a single node and across nodes. Within a single node (small panel in figure~\ref{fig:ubq_scaling}) it is interesting to note that we observe a super linear speedup when going from 1 to 2 or 4 cores, while for more than 8 cores the additional speedup is rather small. We attribute this to the cores' competition for shared resources like cache and memory bandwidth, which become available as more subunits of the machine (e.g.\ both sockets) are used, but are ultimately exhausted when too many processor cores compete for them. The visible oscillations in figure~\ref{fig:ubq_scaling} are due to the use of ScaLAPACK\cite{SCALAPACKUserGuide} to implement equation~\eqref{eq:precalc}, which favors even and especially power of two processor grid sizes.

For on-the-fly calculated atomic transition charges the overall performance is worse, but the parallel scaling both within a node and across nodes is better. Within a single node this is due to the absence of the large matrix of atomic transition charges, which reduces reading from main memory and thereby frees shared resources. The better scaling across nodes is simply due to the fact that a relatively large amount of time is spent on the trivially parallel task of recalculating atomic transition charges, for which no communication is required.

\section{Conclusion}\label{s:Conclusion}

In summary, we have shown that the computational cost of absorption spectra calculations using time-dependent density functional tight binding (TD-DFTB) can be significantly reduced by not considering single orbital transitions with small oscillator strengths. We have found that small selection thresholds do not noticeably affect the accuracy of the result, while already providing sizable computational benefits. This is especially true if the low energy part of the absorption spectrum contains a large number of spurious low-intensity charge-transfer excitations which can be removed through intensity selection, making otherwise infeasible calculations possible. As an example, we have calculated the absorption spectrum of ubiquitin and the spectral shift upon embedding its tyrosine chromophore into the protein environment, and have demonstrated that the accuracy of intensity-selected TD-DFTB is on a par with competing QM/MM methods, which tend to require more work and expertise from the user.

We believe that its ease of use together with the moderate computational cost of intensity-selected TD-DFTB lower the barrier of performing optical properties calculations of large molecules, and can serve to make such calculations possible in a wider array of applications. Intensity-selected TD-DFTB has been implemented in the 2014 release of the ADF molecular modeling suite~\cite{ADF2001}.

\begin{acknowledgement}
The authors thank Onno Meijers for his work on the eigensolver backend. The research leading to these results has received funding from the European Union's Seventh Framework Programme (FP7-PEOPLE-2012-ITN) under project PROPAGATE, Ref.~316897.
\end{acknowledgement}

\begin{suppinfo}
We provide the molecular geometries for all example systems from section~\ref{s:Examples}.
\end{suppinfo}

\bibliography{paper1,literature}

\onecolumn\captionsetup{font=normalsize}

\begin{table}[p]
\caption{\label{tab:scaling_time}Computational complexity of operations within the TD-DFTB method.}
\begin{tabular}{c|c|c|c}
\multirow{2}{*}{Operation} & \multicolumn{3}{c}{Computational complexity of operations per method:} \\
 & direct diag. & Davidson (precalc) & Davidson (on-the-fly) \\ \hline
\hline direct diag. of full~$\mat \Omega$ & $N_\mathrm{trans}^3$ & -- & -- \\
\hline subspace basis orthon. & -- & \multicolumn{2}{c}{$N_\mathrm{trans} N_\text{excit}^2$} \\
\hline mat.-vec. multiplication & -- & \specialcell{$N_\mathrm{trans} N_\mathrm{atom} N_\mathrm{excit}$\\{\small with a small prefactor}} & \specialcell{$N_\mathrm{trans} N_\mathrm{atom} N_\mathrm{excit}$\\{\small with a large prefactor}}
\end{tabular}
\vspace{1.5\baselineskip}\hrule
\end{table}

\begin{table}[p]
\caption{\label{tab:scaling_memory}Memory requirements of the TD-DFTB method. Note that this table only contains the largest objects needed during the diagonalization itself.}
\begin{tabular}{c|c|c|c}
\multirow{2}{*}{Object} & \multicolumn{3}{c}{Scaling of storage requirements per method:} \\
 & direct diag. & Davidson (precalc) & Davidson (on-the-fly) \\ \hline
\hline full matrix~$\mat \Omega$ & $N_\mathrm{trans}^2$ & -- & -- \\
\hline eigenvectors~$\vec F_I$ & $N_\mathrm{trans}^2$ & \multicolumn{2}{c}{$N_\mathrm{trans} N_\text{excit}$} \\
\hline subspace basis vectors & -- & \multicolumn{2}{c}{$N_\mathrm{trans}$ {\small with a large prefactor}} \\
\hline parameter matrix~$\mat \gamma$ & -- & \multicolumn{2}{c}{$N_\mathrm{atom}^2$} \\
\hline atomic transition charges~$\mat h$ & -- & $N_\mathrm{trans} N_\mathrm{atom}$ & -- \\
\hline coefficient matrix~$\mat c$ & -- & -- & $N_\mathrm{atom}^2$ \\
\hline product matrix~$\mat \Theta = \mat S \mat c$ & -- & -- & $N_\mathrm{atom}^2$
\end{tabular}
\vspace{1.5\baselineskip}\hrule
\end{table}

\begin{table}[p]
\caption{\label{tab:timings} Measured runtimes of the example TD-DFTB calculations using intensity selection. The calculations for C$_{60}$ and Ir(ppy)$_3$ were performed on a workstation with an Intel Core i7-4770 processor and 16GB memory. The ubiquitin calculations were performed on 1 to 8 cluster nodes with two octa-core Intel Xeon E5-2650~v2 processors each and 64GB of memory per node.}
\begin{tabular}{c|c|c|c|c|c|c|c}
System & $N_\mathrm{atom}$ & $f_{ia}^\mathrm{min}$ & $N_\mathrm{trans}$ & $N_\mathrm{excit}$ & \#{\small CPU} & $t_\mathrm{wall}$ & $t_\mathrm{CPU}$ \\ \hline
\hline C$_{60}$ & 60 & -- & \multicolumn{2}{c|}{14400} & 4 & 434s & 1736s \\
\hline C$_{60}$ & 60 & 0.001 & \multicolumn{2}{c|}{3610} & 4 & 12s & 49s \\
\hline C$_{60}$ & 60 & 0.005 & \multicolumn{2}{c|}{2581} & 4 & 5.5s & 22s \\
\hline C$_{60}$ & 60 & 0.01 & \multicolumn{2}{c|}{2113} & 4 & 3.5s & 14s \\
\hline C$_{60}$ & 60 & 0.05 & \multicolumn{2}{c|}{1032} & 4 & 0.8s & 3.2s \\
\hline Ir(ppy)$_3$ & 61 & -- & \multicolumn{2}{c|}{7830} & 4 & 88s & 352s \\
\hline Ir(ppy)$_3$ & 61 & 0.001 & \multicolumn{2}{c|}{5656} & 4 & 37s & 148s \\
\hline Ir(ppy)$_3$ & 61 & 0.01 & \multicolumn{2}{c|}{3326} & 4 & 10s & 40s \\
\hline Ir(ppy)$_3$ & 61 & 0.02 & \multicolumn{2}{c|}{2426} & 4 & 4.9s & 20s \\
\hline Ir(ppy)$_3$ & 61 & 0.03 & \multicolumn{2}{c|}{1896} & 4 & 2.8s & 11s \\
\hline Ubiquitin & 1231 & -- & $2\,284\,880$ & 15820 & 128 & 12.6h & 67d \\
\hline Ubiquitin & 1231 & 0.001 & $1\,628\,370$ & 1638 & 48 & 1.1h & 2.1d \\
\hline Ubiquitin & 1231 & 0.02 & $689\,208$ & 552 & 32 & 635s & 5.6h \\
\hline Ubiquitin & 1231 & 0.05 & $333\,337$ & 359 & 16 & 332s & 1.5h \\
\hline Ubiquitin & 1231 & 0.1 & $156\,488$ & 232 & 16 & 137s & 2192s
\end{tabular}
\end{table}

\end{document}